\documentclass[11pt,a4j]{article}
\usepackage[dvips]{graphicx}
\usepackage{mathrsfs}
\usepackage{amsthm} 
\usepackage{amsmath}
\usepackage{amssymb}
\usepackage{amsfonts}
\usepackage{yhmath}
\usepackage{wrapfig}
\usepackage[dvips]{color}
\usepackage{cite}
\usepackage{array}
\usepackage{tabularx}
\usepackage[sectionbib]{chapterbib}
\usepackage{threeparttable}
\usepackage{chemarrow}
\usepackage{ascmac}
\usepackage{framed}
\definecolor{shadecolor}{gray}{0.90}

\begin{document}

\renewcommand{\figurename}{\small{Fig.}~}
\renewcommand{\labelitemi}{}
\renewcommand{\thefootnote}{$\dagger$\arabic{footnote}}
\renewcommand{\footnoterule}{%
  \vspace{2pt}
\flushleft\rule{6.154cm}{0.4pt}
  \vspace{4pt}
}
\pagestyle{plain}

\begin{flushright}
\textit{Gelation Paradox}
\end{flushright}
\vspace{0mm}

\begin{center}
\setlength{\baselineskip}{20pt}{\LARGE\textbf{Gelation Paradox}}
\end{center}
\vspace{-2mm}

\begin{center}
\large{Kazumi Suematsu} \vspace*{2mm}\\
\normalsize{\setlength{\baselineskip}{14pt} 
Institute of Mathematical Science\\
Ohkadai 2-31-9, Yokkaichi, Mie 512-1216, JAPAN\\
E-mail: suematsu@m3.cty-net.ne.jp,  Tel/Fax: +81 (0) 593 26 8052}\\[6mm]
\end{center}

\hrule
\vspace{0mm}
\begin{flushleft}
\textbf{\large Abstract}
\end{flushleft}
\vspace{-2mm}
\setlength{\baselineskip}{14pt}
The gelation paradox first raised by Stockmayer is re-examined by comparing the extent of reaction in sol phase and that in the interior of gel. It is shown that the Stockmayer limit, $2/f$, for the extent of reaction is the lowest limit in gel phase, as well as being the highest limit for sol molecules.
\\[-5mm]
\begin{flushleft}
\textbf{\textbf{Key Words}}:
\normalsize{Extent of Reaction/ Interior of Gel/ Infinity}\\[3mm]
\end{flushleft}
\hrule
\vspace{5mm}

In this paper, we discuss the gelation paradox. The question was first raised by Stockmayer in 1949\cite{Stockmayer5}, but formally has not been answered until today. The question associated with the paradox is, according to Stockmayer, as follows:\\[-3mm]

\textit{Consider the ideal branching process of the R$-$A$_{f}$ model. Assume the equal reactivity of functional units, but that no ring formation is allowed. Consider an $x$-mer molecule. This molecule has $f\hspace{0.2mm}x$ functional units and $2(x-1)$ reacted functional units, so that the extent of reaction of this molecule is $2(x-1)/f\hspace{0.1mm}x$. Since gel is an infinitely large molecule, the extent of reaction in the interior of gel must be
\begin{eqnarray}
p_{gel}=\lim_{x \to \infty}\frac{2(x-1)}{f\hspace{0.1mm}x}\rightarrow\frac{2}{f}\label{gparadox-1}
\end{eqnarray}
It is apparent that the extent of reaction of the whole system can not exceeds $2/f$ because this is the upper limit for the gel phase to attain. Quite in contrast, according to the Flory formulation\cite{Flory5}, the system easily exceeds this point and attains $p=1$; hence one must conclude that the excess bonds beyond $2/f$ are due to ring formation. Quite aside from the fundamental logic involved in Flory's procedure, his formulation cannot be rigorous, for it predicts a very definite number of cyclic linkages in the system, although the initial assumptions do not specifically treat such structures in any way.}\\[-3mm]

The above argument by Stockmayer reveals that the ideal branching theory that forbids ring formation in its initial assumption allows itself the occurrence of rings\cite{Gordon, Erukhimovich}; it shows that the classic theory includes the paradoxical logic that the conclusion contradicts the initial assumption. A purpose of this short paper is to give another explanation for the old question from a different point of view, since the question has revived repeatedly in the community, formally and informally.

\section*{Extent of Reaction in the Interior of Gel}
Let us confine our discussion to the regime: $p_{c}\le p\le1$. The extent of reaction of sol and gel can be defined by
\begin{equation}
p_{sol}=\frac{\displaystyle 2\sum\nolimits_{x=1}^{\infty}(x-1)N_{x}}{\displaystyle f\sum\nolimits_{x=1}^{\infty}xN_{x}}\label{gparadox-2}
\end{equation}
\begin{equation}
p_{gel}=\frac{\displaystyle fM_{0}p-2\sum\nolimits_{x=1}^{\infty}(x-1)N_{x}}{\displaystyle f\left(M_{0}-\sum\nolimits_{x=1}^{\infty}xN_{x}\right)}\label{gparadox-3}
\end{equation}
It is known that the number distribution of $x$-clusters has the form:
\begin{equation}
N_{x}=M_{0}\frac{f\{(f-1)x\}!}{x! \,\nu_{x}!}\,p^{x-1}(1-p)^{\nu_{x}}\hspace{5mm}(x=1, 2, \cdots)\label{gparadox-4}
\end{equation}
where $\nu_{x}=(f-2)x+2$ represents the number of unreacted functional units on an $x$-cluster. Let $\theta=p\,(1-p)^{f-2}$, then define the generating function:
\begin{equation}
A(\theta(p))=\sum_{x=1}^{\infty}N_{x}/M_{0}=\frac{(1-p)^{2}}{p}\sum_{x=1}^{\infty}\omega_{x}\theta^{x}\label{gparadox-5}
\end{equation}
where $\omega_{x}=\frac{f\{(f-1)x\}!}{x! \,\nu_{x}!}$. The weight fraction of sol is, by definition,
\begin{equation}
W_{sol}=\sum_{x=1}^{\infty}xN_{x}/M_{0}=\frac{(1-p)^{2}}{p}\sum_{x=1}^{\infty}x\,\omega_{x}\theta^{x}\label{gparadox-6}
\end{equation}
Our aim is to calculate the quantity $A(\theta)$. For this purpose, define the other generating function:
\begin{equation}
\theta\frac{\partial M(\theta)}{\partial\theta}=\sum_{x=1}^{\infty}x\,\omega_{x}\theta^{x}\label{gparadox-7}
\end{equation}
from which we have
\begin{equation}
M(\theta)=\sum_{x=1}^{\infty}\omega_{x}\theta^{x}+c_{1}\label{gparadox-8}
\end{equation}
In order to equate $M(\theta)$ with $\sum_{x=1}^{\infty}\omega_{x}\theta^{x}$, we put $c_{1}=0$. The boundary condition then becomes
\begin{equation}
M(\theta)\mid_{\theta=0}=M(\theta(p))\mid_{p=1}=0\label{gparadox-9}
\end{equation}

\section*{Solution for a Special Case of $f=3$}
\begin{figure}[h]
\vspace*{4mm}
\begin{center}
\includegraphics[width=8.5cm]{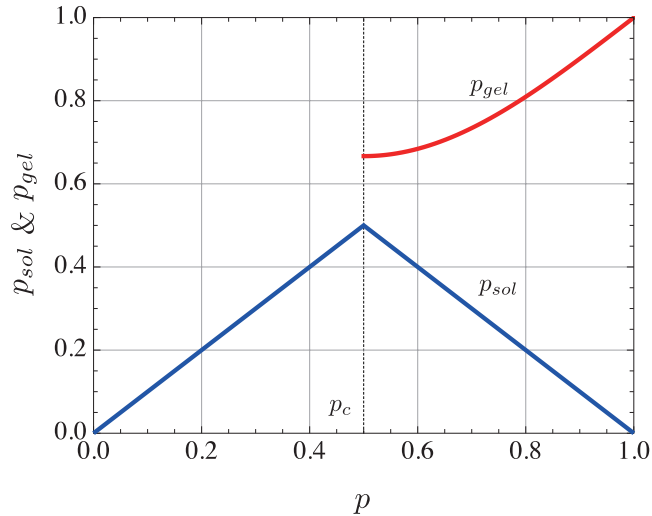}
\caption{The extent of reaction of sol and gel. $p_{sol}$ is the theoretical line by eq. (\ref{gparadox-18}) and $p_{gel}$ is that by eq. (\ref{gparadox-19}).}\label{Gel-Paradox}
\end{center}
\end{figure}

Before proceeding with our discussion, let us consider the familiar problem\cite{Essam5}: Let $Q$ be the probability that a chosen branch is finite. Then the following recurrence relation holds:
\begin{equation}
Q=1-p+p\,Q^{f-1}\label{gparadox-10}
\end{equation}
which has the solutions: $Q=1$ and
\begin{equation}
p\left(Q^{f-2}+Q^{f-3}+\cdots+1\right)=1\label{gparadox-11}
\end{equation}
For $f=3$, eq. (\ref{gparadox-11}) yields
\begin{equation}
Q=\frac{1-p}{p}\label{gparadox-12}
\end{equation}
from which we have
\begin{equation}
W_{sol}=\sum_{x=1}^{\infty}xN_{x}/M_{0}=\left(\frac{1-p}{p}\right)^{3}\hspace{5mm}\text{for }\,p\ge p_{c}\label{gparadox-13}
\end{equation}
Then using eqs. (\ref{gparadox-6}) and (\ref{gparadox-7}), we have
\begin{equation}
\theta\frac{\partial M(\theta)}{\partial\theta}=\frac{1-p}{p^{2}}\label{gparadox-14}
\end{equation}
so that, by $\theta=p(1-p)$ for $f=3$, we have
\begin{equation}
\frac{\partial M(\theta)}{\partial\theta}=p^{-3}\label{gparadox-15}
\end{equation}
Since $d\theta/dp=1-2p$,
\begin{equation}
M(\theta(p))=\sum_{x=1}^{\infty}\omega_{x}\theta^{x}=-\frac{1}{2p^{2}}+\frac{2}{p}+c_{2}\label{gparadox-16}
\end{equation}
Using  the boundary condition, $M(p)\mid_{p=1}=0$, we have $c_{2}=-3/2$. Then substituting eq. (\ref{gparadox-16}) into eq. (\ref{gparadox-5}), we gain
\begin{equation}
A(\theta(p))=\sum_{x=1}^{\infty}N_{x}/M_{0}=\left(\frac{1-p}{p}\right)^{3}\frac{3p-1}{2}\hspace{5mm}\text{for }\,p\ge p_{c}\label{gparadox-17}
\end{equation}
By eqs. (\ref{gparadox-2}) and (\ref{gparadox-3}) along with eqs. (\ref{gparadox-13}) and (\ref{gparadox-17}), we can now express $p_{sol}$ and $p_{gel}$ as functions of $p$; the result is
\begin{align}
p_{sol}=&1-p\hspace{5mm}\text{for }\,p\ge p_{c}\label{gparadox-18}\\[3mm]
p_{gel}=&\frac{\displaystyle p\left\{1-\left(\frac{1-p}{p}\right)^{4}\right\}}{\displaystyle\left\{1-\left(\frac{1-p}{p}\right)^{3}\right\}}\hspace{5mm}\text{for }\,p\ge p_{c}\label{gparadox-19}
\end{align}
The above solutions are plotted in Fig. \ref{Gel-Paradox}. it is seen that while $p_{sol}$ decreases linearly from 1/2 at $p=p_{c}$ to 0 at $p=1$, $p_{gel}$ increases, starting from $p_{gel}=2/3=2/f$ in question, monotonically to 1 at $p=1$, quite in contrast to the equality given in eq. (\ref{gparadox-1}). Stockmayer argues that the excess bonds beyond $p_{c}$ must be ascribed to ring formation within the gel phase.

It is important to notice, however, that when we have the argument on eq. (\ref{gparadox-1}), we are talking about sol, because the equality $2(x-1)/f\hspace{0.1mm}x$ itself assumes that the molecule has $(f-2)x+2$ unreacted functional units, namely a finite molecule\cite{Kazumi5}. This reveals that the quantity $p=2/f$ is the upper limit of $p_{sol}$, but not that of $p_{gel}$. It is the limiting point a largest sol molecule can attain.

\begin{figure}[h]
\vspace*{4mm}
\begin{center}
\includegraphics[width=9cm]{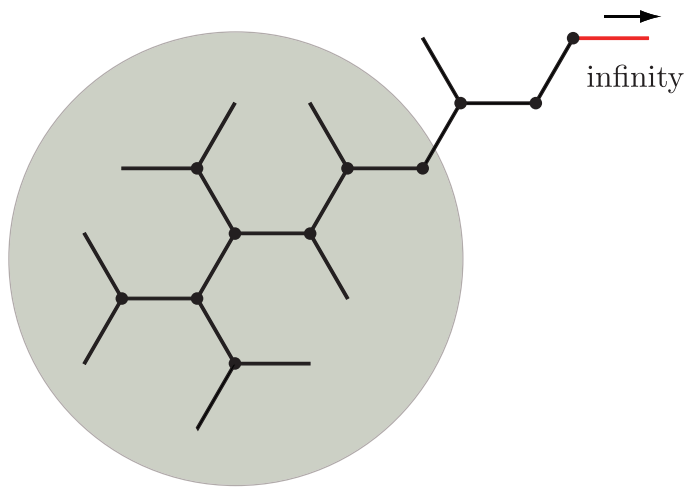}
\caption{An example of gel molecule. Note that at least one end must lead to \textit{infinity}.}\label{Gel-Molecule}
\end{center}
\end{figure}

The above argument becomes more apparent by making reference to Fig. \ref{Gel-Molecule}. Since gel is an infinite molecule, it must possess, at least, one end that leads to infinity. This end must be counted as a reacted functional unit. In other words, the total number of the reacted functional units within gel must be larger than $2(x-1)+1$. Mathematically
\begin{equation}
p_{gel}\ge\lim_{x \to \infty}\frac{2(x-1)+1}{f\hspace{0.1mm}x}=\frac{2}{f}\label{gparadox-20}
\end{equation}
We realize that $p=2/f$ is the lowest limit of $p_{gel}$, as well as being the highest limit for sol molecules. In mathematical terms it must be that
\begin{equation}
0\le p_{sol}^{1}\le 2/f\le p_{gel}\le 1\label{gparadox-21}
\end{equation}
where we use the notation $p_{sol}^{1}$ to express the extent of reaction for a single sol molecule and to distinguish from that, $p_{sol}$, for the whole sol phase. In contrast to the Stockmayer argument, the extent of reaction in the interior of gel must be greater than $2/f$, in accord with the result of Fig. \ref{Gel-Paradox}. In the mathematical point of view, therefore, there is no paradox in the Flory formulation.

\end{document}